%% file: 0_main.tex
\documentclass{article}
\usepackage{ijcai21}

\usepackage{times}
\usepackage{soul}
\usepackage{url}
\usepackage[hidelinks]{hyperref}
\usepackage[utf8]{inputenc}
\usepackage[small]{caption}
\usepackage{graphicx}
\usepackage{amsmath}
\usepackage{amsthm}
\usepackage{booktabs}
\usepackage{algorithm}
\usepackage{algorithmic}
\usepackage{color}
\usepackage{comment}
\usepackage{graphicx}

\usepackage{amssymb}% http://ctan.org/pkg/amssymb
\usepackage{pifont}% http://ctan.org/pkg/pifont
\newcommand{\cmark}{\ding{51}}%
\newcommand{\xmark}{\ding{55}}%
\usepackage{subcaption}
\graphicspath{{figs/}} 
\newcommand{\sys}{TAD}

\title{\sys{}: Trigger Approximation based Black-box Trojan Detection for AI}
\author{
Xinqiao Zhang$^1$\footnote{Contact Author}\and
Huili Chen$^1$\And
Farinaz Koushanfar$^1$\\
\affiliations
$^1$UC San Digeo\\
\emails
\{x5zhang, huc044, fkoushanfar\}@ucsd.edu
}

\begin{document}
% \verb|\checkmark|: \checkmark \par
% \verb|\cmark|: \cmark \par
% \verb|\xmark|: \xmark

\maketitle

\vspace{-1em}
\begin{abstract}
An emerging amount of intelligent applications have been developed with the surge of Machine Learning (ML). Deep Neural Networks (DNNs) have demonstrated unprecedented performance across various fields such as medical diagnosis and autonomous driving. 
While DNNs are widely employed in security-sensitive fields, they are identified to be vulnerable to Neural Trojan (NT) attacks that are controlled and activated by the stealthy trigger. We call this vulnerable model as adversarial artificial intelligence (AI).
In this paper, we target to design a robust Trojan detection scheme that inspects whether a pre-trained AI model has been Trojaned before its deployment. 
Prior works are oblivious of the intrinsic property of trigger distribution and try to reconstruct the trigger pattern using simple heuristics, i.e., stimulating the given model to incorrect outputs. As a result, their detection time and effectiveness are limited.  
We leverage the observation that the pixel trigger typically features spatial dependency and propose \sys{},
the first trigger approximation based Trojan detection framework that enables fast and scalable search of the trigger in the input space. 
Furthermore, \sys{} can also detect Trojans embedded in the feature space where certain filter transformations are used to activate the Trojan. 
We perform extensive experiments to investigate the performance of the \sys{} across various datasets and ML models. 
Empirical results show that \sys{} achieves a ROC-AUC score of $0.91$ on the public TrojAI dataset~\footnote{https://pages.nist.gov/trojai/docs/data.html} and the average detection time per model is $7.1$ minutes. 
\end{abstract}

\input{1_intro}
\input{2_background}

\input{3_overview}
\input{4_methodsummary}

\input{5_evaluation}
\input{6_conclusion}

\section*{Acknowledgments}
This work is supported by Intelligence Advanced Research Projects Activity(IARPA). The project number is W911NF20C0045. And this work is partially funded by SRC auto 2899.001.

\begin{comment}
\section*{Acknowledgments}
The was was supported by International Research and Publishing Academy(IRAPA). The project number is W911NF20C0045.
\end{comment}

\bibliographystyle{unsrt}
\bibliography{./ref}

\end{document}

%% file: 1_intro.tex
\vspace{-0.8em}
\section{Introduction}

Artificial intelligence (AI) has been widely investigated and offers tremendous help in different fields such as image classification, speech recognition, and data forecasting~\cite{huang2019real}. With the help of AI, some applications such as self-driving cars~\cite{rao2018deep}, spam filter~\cite{nosseir2013intelligent} and face recognition~\cite{sun2015deepid3} offer many conveniences in human's lives.

AI also brings the potential risk with the application. Trojan attacks, also called backdoor attacks, aim to modify the input of the data by adding a specific trigger to make the AI output the incorrect response~\cite{karra2020trojai}. The trigger data is very rare in the dataset so that usually it is impossible to be aware of. Once the trigger has been applied to the input data, the output result will be totally different than the result that was supposed to be, which will cause big security issues. No one wants these issues to happen when in a self-driving car. The emerging concerns of security have brought dramatic attention to the research domain.
Therefore, a fast and accurate detection method needs to be introduced to detect if a given model has been trojaned or backdoored by a hacker. These concerns lead us to the idea of this paper. We introduce a novel method to detect two popular backdoor trigger attacks-- polygon attacks and Instagram filter attacks.

Polygon attacks are the most popular and most investigated attacks. Adversaries typically add a polygon image of a certain shape on top of many input images and use these perturbed images during the training stage, aiming to mislead the model to produce incorrect responses~\cite{wang2019neural}. 
A model is poisoned or trojaned if it yields unexpected results when the input images contain the pre-trained polygon trigger. Typically, the poisoned images take a small percentage (20\% or less) of the full training data. 
Another Trojan attack, Instagram filter attacks, applies an Instagram filter to a given image and the output will be completely different from the original class. 
Compared to polygon triggers that directly replace the pixels of a certain region, the Instagram filter triggers transform the whole image (i.e., change the color value of every pixel) based on the filter type to produce the poisoned images~\cite{Trojai}.

Prior works focus on different ways of detection, either black-box detection DeepInspect (DI)~\cite{chen2019deepinspect} or white-box detection Neural Cleans (NC)~\cite{wang2019neural}.
DI learns the probability distribution of potential triggers from the queried model using a conditional generative model. The good side is that it does not require a benign dataset to assist backdoor detection, the drawback is that the running time is very long. NC is one of the first robust and generalizable method that targets backdoor detection and mitigation. 
The proposed method can identify the backdoored model and reconstruct possible triggers from the given model. However, the limitation is that the scalability is not very good and their technique only applies to a single model and a single dataset. Also, it takes time to fine-tune the hyper-parameters when switching to another model and dataset.  
In real life, we usually have a number of models with different architectures running at the same time for different jobs. Given this information, the Neural Cleans needs to be improved.
Another work TABOR~\cite{guo2019tabor} proposes an accurate approach to inspect Trojan backdoors. The paper introduces a metric to quantify the probability that the given model contains the Trojan.
TABOR achieves a feasible performance but the method is evaluated on a limited number of datasets and DNN architectures. Moreover, the trigger is attached at a pre-known location before detection. As such, TABOR is limited for practical application.

To the best of our knowledge, \sys{} is the first Trigger Approximation based Black-box Trojan Detection method for DNN security assurance. \sys{} takes a given model as a black-box oracle and characterizes the Trojan trigger by several key factors % and leverages some dominate parameters 
to detect if the model is poisoned or not. It combines the advantages of both DI and NC while alleviating their drawbacks. 
Our method achieves a high detection performance, robustness, and low runtime overhead. 
Moreover, the proposed method can detect random triggers attached to any location in the foreground image for DNN models across a large variety of architectures, which is a big improvement compared to TABOR~\cite{guo2019tabor}.

In summary, our contributions are shown as follows:
\begin{itemize}
% \squishlist
    \vspace{-0.3em}
    \item \textbf{Presenting the first robust and fast method for random trigger Backdoor detection.} Our approximate trigger reconstruction method facilitates efficient exploration of a given model, thus is more robust and fast compared to the existing approach.
    
    \vspace{-0.2em}
    \item \textbf{Enabling both polygon trigger detection and Instagram filter detection.} \sys{} provisions the capability of detecting the most common used triggers and yielding an estimation of poison probability.

    \vspace{-0.2em}
    \item \textbf{Investigating the performance of \sys{} on diverse datasets and model architectures.} We perform an extensive assessment of \sys{} and show its superior effectiveness, efficiency, and scalability.  

\end{itemize}

%% file: 2_background.tex
\vspace{-1em}
\section{Background}

% \subsection{Data training}
\subsection{Backdoor Attacks}

Backdoor attacks are typically implemented by data poisoning and may have different objectives, e.g. targeted attacks, untargeted attacks, universal attacks and constrained attacks. 
The source class is the class that clean image comes from. 
Generally, a backdoored image consists of a clean image and a trigger. 
A trigger can be an image block or a filter transformation that is added to the clean image.
The attack target class is the output class of the poisoned model on the backdoored image. The target class can be any class other than the source class of the image. In this paper, we assume a poisoned model has a single target class.

Data poisoning attacks are where users use false data during the training process, leading to the corruption of a learned model~\cite{steinhardt2017certified}. This kind of attack is very popular and commonly used for backdoor attacks. Usually, it is not clear to find a model has been attacked or not because the corrupting response of a model shows only when partial false data exists in the input image. It brings a very big risk for some applications like self-driving cars and it may cause an undesirable consequence. Therefore, data poisoning attacks are an emerging issue nowadays.

\noindent \textbf{(i) Targeted attacks vs untargeted attacks.} Trojan attacks can be categorized into two types based on the attack objective.
On the one hand, \textit{targeted attack} aims to mislead the model to predict a pre-specified class when input data has certain properties~\cite{Trojannn}. 
Basically, data poisoning attacks do not assign a specific class to the poisoned data. An attacker can use a random class or let the neural network choose the closest target class for the triggered data. 
Targeted attacks are powerful since they enforce neural networks to produce a pre-defined target class while preserving high accuracy on the normal data~\cite{Trojannn}.
On the other hand, \textit{untargeted Trojan attack} aims to degrade the task accuracy of the model on all classes~\cite{gu2017badnets}.

\noindent \textbf{(ii) Universal attacks vs constrained attacks.}
% \textcolor{blue}{all input classes poisoned to the specific target class. }
Trojan attacks can also be categorized by their impact range on the input data.  
(i) \textit{Universal attacks}~\cite{moosavi2017universal} refer to universal(image-agnostic) perturbation that applies to the whole image and all source classes of the model, which leads to the misconceiving of one model. 
Let's assume we have a single small image perturbation that can let the latest DNNs yield incorrect responses, and the small image perturbation can be a vector, a filter or something else.  
Usually, once a universal perturbation is added to an image, the target class will always be the same one.
This kind of perturbation has been investigated in ~\cite{moosavi2017universal} and the paper~\cite{moosavi2017universal} introduces an algorithm to find such perturbations. Usually, a very small perturbation can cause an image to be misclassified with a very high probability. The basic idea is to use a systematic algorithm for computing universal perturbations. They find that these universal perturbations generalize very well across neural networks, bring potential security breaches. 

% How does it work?
\textit{Constrained attacks} refer to the trigger that is only valid for the pre-defined source classes. It leads to a poisoned model that has source classes which are a subset of all classes. In this case, only images in source classes are poisoned during the training process.

\noindent \textbf{(iii) Polygon triggers and Instagram filter triggers}
There are two types of Trojan triggers based on their insertion space: 
polygon triggers and Instagram filter triggers. 
(i) Polygon triggers are inserted in the \textit{input space}, superimposed directly on the original benign images. More specifically, the polygon trigger has a specific shape (also called trigger mask) and color, added to the image at one location.
The color of a polygon can be any value from $0$ to $255$ for each channel.
(ii) Instagram triggers refer to combinations of image transformations that are applied on the clean images.
Note that both polygon and Instagram triggers can be used for targeted/untargeted and universal/constrained attacks.

% \vspace{-1.1em}
\begin{figure}[ht!]
    \setlength{\belowcaptionskip}{-10pt}
    \centering
    \includegraphics[width=0.4\textwidth]{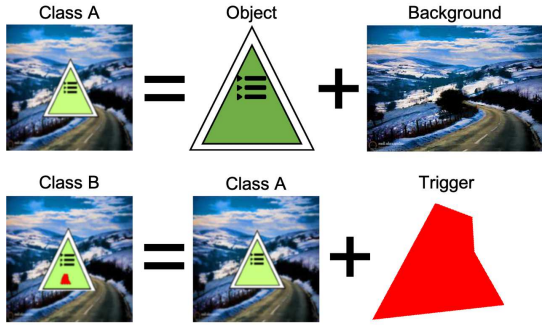} 
    \vspace{-1.3em}
    \caption{Example of clean image and polygon trigger~\protect\cite{Trojai}}
    \label{fig:example}
    % \vspace{-2em}
\end{figure}

\begin{figure}[ht!]
    \centering
    \includegraphics[width=0.42\textwidth,scale=0.2]{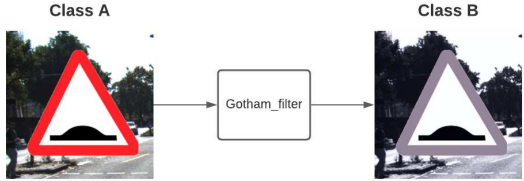} 
    \vspace{-1.3em}
    \caption{Example of clean image and Instagram filter trigger}
    \label{fig:example_ins}
    \vspace{-2em}
\end{figure}

Figure~\ref{fig:example} and Figure~\ref{fig:example_ins} show the example of triggers being implemented into clean images. Generally, for polygon trigger, a clean image consists of one foreground object image and one background image while the poison image is made from a clean image by adding a scaled trigger image at a specific location. The Instagram filter trigger applies to the complete image.

\vspace{-1.0em}

\subsection{Backdoor Detection}
 
% \huc{prior works (more details than introduction). }
Prior works apply different methods for backdoor detection. Generally, model-level detection and data-level detection are two common kinds of detection.
Neural Cleanse(NC)~\cite{wang2019neural} explores model-level detection, NC treats a given label as a potential victim class of a targeted backdoor attack. Then NC designs an optimization scheme to find the minimal trigger required to misclassify all samples from source classes into the victim class. After several rounds, NC finds some potential triggers. During the experiment, an outlier detection method is used to choose the smallest trigger among all potential triggers. A significant outlier represents a real trigger and victim class is the target label of the backdoor attack. The drawback is that NC only takes some certain trigger candidates into consideration and it does not work well for random polygon triggers.

% \noindent \textbf{(i) model-level detection: } 
DeepInspect(DI)~\cite{chen2019deepinspect} is another model-level detection method. DI is the first black-box Trojan detection solution with limited prior knowledge of the model. DI first learns the probability distribution of potential triggers from the given models using a conditional generative model and retrieves the footprint of backdoor insertion. DI has a good detection performance and a smaller running time compared to prior work. The downside of DI is that the runtime increases much for large models, which probably exceeds the timing requirements.

ABS~\cite{liu2019abs} is also a model-level detection method. ABS uses a technique to scan neural network based AI models to determine if a given model is poisoned. ABS analyzes inner neuron behaviors by determining how output activations change when we introduce different levels of stimulation to a neuron. A substantially elevating the activation of a specific output label in a neuron regardless of the input is considered potentially poisoned. Then a reverse engineering method is applied to verify that the model is truly poisoned. Even though ABS performs extensive experiments, the accuracy is a big issue when the number of neurons of a certain layer is increasing because sweeping each neuron requires to pick a step size carefully. Thus it is not guaranteed to have a good performance for large AI such as $densenet121$.

For data-level detection, CLEANN~\cite{javaheripi2020cleann} is a first end-to-end framework and very lightweight methodology which recovers the ground-truth class of poison samples without the requirement for labeled data, model retraining, or prior assumptions on the trigger or the attacks. By applying sparse recovery and outlier detection, it can detect the attack at the data-level. However, it can only detect specific types of triggers e.g, square, Firefox and watermark.

%% file: 3_overview.tex
\vspace{-1em}
\section{Overview}
\vspace{-0.3em}
In this section, we present the overview of our paper. The following part talks about the method we proposed, and then we discuss the threat model. After that, we give detailed information about our algorithms. Then, we list our experiment results and analysis.

\vspace{-0.4em}
\subsection{Threat model}

Our method examines the susceptibility of the given AI with very limited assumptions. We assume we have no access to the clean image that used for training.
Also, we consider two types of triggers, polygon triggers and Instagram filter triggers. The potential candidates of Instagram filters are given. For polygon triggers, we do not assume prior knowledge about the trigger information e.g, trigger location, trigger size, trigger sides and shape, trigger color. Trigger location can be any location inside of the foreground image and size ranges from 0 to 0.25. Trigger sides, shape and color are all random.

\subsection{\sys{} Global Flow}

\vspace{-1.1em}
\begin{figure}[ht!]
    \centering
    \includegraphics[width=0.4\textwidth,scale=0.2]{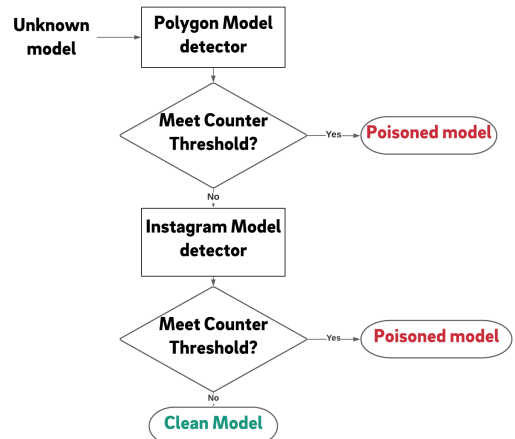} 
    \vspace{-1.3em}
    \caption{Flow of Trojaned AI detection method. }
    \label{fig:flow}
    \vspace{-1em}
\end{figure}

Figure~\ref{fig:flow} shows the overall flow of our method, Our detection method has two sequential stages. Given an unknown model, we first check if it is backdoored by polygon Trojans (S1). If the detection result of this stage is benign, we then check if the model is backdoored by Instagram filter Trojans (S2). Only models that pass both detection stages are considered benign models. The order of two stages is important because we find out that some Instagram poisoned models can be triggered by a polygon image as well.

\textbf{(S1)} To detect polygon poisoned models, we recover the trigger for each source class. We try different trigger parameters and see if the predictions of the model have a high bias towards a specific class. If this prediction bias is observed, then the model is determined as Trojaned. 

\textbf{(S2)} To detect Instagram filter poisoned models, we search for each (source, target) class pair and apply each of the five filter types individually. If the images in the source class have a high probability to be predicted as the target class after applying a certain filter, then the model is decided as Trojaned. 

%% file: 4_methodsummary.tex
\vspace{-0.4em}
\section{Methodology}
\vspace{-0.2em}

\subsection{Preliminary Experiments and Observations}

\noindent \textbf{Trigger Characterization.}
Our preliminary experiments explore the impacts of different trigger parameters of polygon triggers. 
In our experiment, we mainly use four trigger parameters which are trigger location, trigger shape, trigger color and trigger size. Trigger location is the center location that trigger is attached to a clean image and trigger shape includes the shape of the trigger and the number of sides, note that the the length of each side is a random number. Trigger color is a 3 channel RGB value ranging from 0 to 255, e.g. $[200,0,255]$. A \textit{trigger mask} is a full size of trigger without RGB value and trigger size is the scaling parameter from $0.01$ to $0.25$ applied to a trigger mask.
First we define a generic form of polygon trigger injection:
\begin{equation}
    \vspace{-0.3em}
    T(X,l,v) = X_{eb},
    \vspace{-0.3em}
\end{equation}
where $T(\cdot)$ represents the function that attaches an array of trigger value $v$ to the clean image. $v$ is a 3D matrix of pixel color value that shares the same dimension of the input image such as height, width, and color channels. $l$ is a 2D matrix called trigger mask location. Trigger mask location is the location that the trigger will overwrite the clean image and its value is either $0$ or $1$. $X_eb$ is a trigger embedded image defined as: 
\begin{equation}
\vspace{-0.3em}
X_{eb(i,j,k)}  = \left\{\begin{matrix}
v_{(i,j,k)} + s*X_{(i,j,k)} & if~ l_{i,j} =1\\ 
X_{(i,j,k)} &  Otherwise \\
\end{matrix}\right.
\vspace{-0.1em}
\end{equation}

For a specific pixel location $i,j$ in a image, if $l_{i,j} = 1$, then the value of the pixel will be overwritten to $X_{eb(i,j,k)} = v_{(i,j,k)} + s*X_{(i,j,k)}$, while when $l_{i,j} = 0$, it will keep the original value  $X_{eb(i,j,k)} = X_{(i,j,k)}$. Note that for polygon trigger, the value of mask location $l$ is 0 or 1 and $s$ is 0. For Instagram filter trigger the value $s$ is continuous from 0 to 1 based on the filter configuration.

% \huc{add mathematical formulation for trigger mask and values. see Neural Cleanse. } \xq{add equation} 
First, to get the overall idea of how the polygon trigger impacts the output of a model, we investigate different trigger characterization factors(trigger parameters) individually and get the output. Trojan Activation Rate represents the frequency that an image gets an incorrect response. Then we have the following observations for polygon triggers: 

\textbf{Observation 1: Given the other correct trigger parameters, trigger location does not impact the Trojan Activation Rate.} We find that a Trojan model will get a constant response when applying the trigger to anywhere on the clean image while keeping other trigger parameters correct. We believe it is the 2D convolution layer inside of the model and the trigger will be always detected by these convolutional neurons and output incorrect response. 
% \huc{Is this true when only fixing other factors, or the other factors need to be correct? For the former case, I think it's possible to find an counter example.} \xq{yes, true}

\textbf{Observation 2: Trigger shape has a very limited influence for Trojan Activation Rate} Some trigger shapes can be shared with other models.

\textbf{Observation 3: Trigger size is very important.} Trigger sizes play an important role to determine if the victim class can be found or not. The size should be the exact value in the correct size range.

\textbf{Observation 4: More than one trigger colors share the same Trojan Activation Rate as the correct trigger color offers} For trigger color, we find that random color has over 50\% chance to induce a fair amount of poison response, which is good enough to set up a threshold to classify clean and poisoned models.

For trigger parameter selection, based on the above observations, we search over $trigger~color$ and $trigger~size$ to perform our further experiment and Table~\ref{tab:diff_para} shows the detailed information about the trigger parameter selection approach. The correct mark means it changes the value of the column name while fixing the other values as the correct value. For example, the second row represents it iterates all possible trigger locations and at the same time keep trigger shape, color, size as ground-truth value, then the $Trojan~Activation~Rate$ is always 1.

\vspace{-0.7em}
\begin{table}[ht!]
\centering
\scalebox{0.8}{
\begin{tabular}{lllll}
\hline
\begin{tabular}[c]{@{}l@{}}Trigger \\ location\end{tabular} & \begin{tabular}[c]{@{}l@{}}Trigger \\ shape\end{tabular} & \begin{tabular}[c]{@{}l@{}}Trigger \\ color\end{tabular} & \begin{tabular}[c]{@{}l@{}}Trigger \\ size\end{tabular} & \begin{tabular}[c]{@{}l@{}}Trojan \\ activation rate\end{tabular}  \\ \hline
\cmark & \xmark & \xmark & \xmark & 1  \\ \hline
\xmark & \cmark & \xmark & \xmark & 0.8-1 \\ \hline
\xmark & \xmark & \cmark & \xmark & 0-1  \\ \hline
\xmark & \xmark & \xmark & \cmark & 0-1  \\ \hline
\end{tabular}
}
\vspace{-0.7em}
\caption{\label{tab:diff_para} Metric for different trigger parameters, \cmark = Change and \xmark = Not change.}
\end{table}
\vspace{-1.5em}
% \huc{report this metric for different trigger params.}

%% 
\vspace{-0.3em}
\subsection{Polygon Trojan Detection}
\vspace{-0.3em}
Since trigger shape does not play a much important role in finding the trigger to activate the poisoned model, we try to find a most common shape and make it work on as many poisoned models as possible. We finally approximate the polygon trigger as a square bounding box from around 1000 possible trigger images. For trigger location, we use the center point of the image as the pre-defined location.

Algorithm~\ref{alg:polygon detection} shows our basic flow of our method. First, we need to initialize two counters, $trigger~counter$ and $round~counter$. 
The $trigger~counter$ records the number of triggers that can activate the model from the source class to the possible victim class 
In each round, we randomly choose one combination of RGB values as the trigger color. 
$Round~Counter$ records the total times that different random colors are used. 
Then, given a random color, image generator function $ImgGen$ uses random RGB color, pre-defined center location and possible trigger sizes to generate different sizes trigger images $I$.
After that, since the total classes of each model is not given, we calculate the total number of classes $T$ by feeding in a random image. Next, we iterate each class of the model and apply all the generated trigger image in $I$ to each clean image $Img$. 

If the output class is not the original source class, the trigger counter will increment by 1. Once the trigger counter reaches the max number to inspect the polygon model $(PO_{max})$, the model will be classified as a poisoned model.
% \huc{it's vague to say `very high probability'}. 
Otherwise, our method uses another trigger color to repeat the process. The max number to initialize trigger color is $CO_{max}$ and if \sys{} cannot find a trigger color that flips one image more than $PO_{max}$ times, then the model will not be flagged as a poisoned model. \sys{} then performs Instagram model detection method.

\vspace{-0.5em}
\begin{algorithm}[ht!]
\caption{Polygon Model Detection. }
\label{alg:polygon detection} 
\textbf{Input}: Model file ($ID_n$) which includes both topology and weights; Clean images for each output class ($Img$); .\\
\textbf{Parameter}: Threshold value to classify polygon model ($Th$), Max count to classify polygon model (${PO}_{max}$); Max rounds to initialize trigger color (${CO}_{max})$. Pre-defined trigger center location ($l_x,l_y$), possible trigger sizes ($S$)\\
% \textbf{Parameter}: Optional list of parameters\\
\textbf{Output}: A single probability that the model was poisoned ($P_1 = high~probability, P_2 = low~probability$).

\begin{algorithmic}[1]
% \vspace{0.2em}
\STATE Load model: $ID_n$.

\STATE Initialize trigger counter: $Trigger~Counter$
% \vspace{0.2em}
\STATE Initialize round counter: $Round~Counter$ 
% \vspace{0.2em}
\STATE Initialize trigger color: $C_1 \gets (R_1,G_1,B_1)$
% \vspace{0.2em}
\STATE Generate trigger image: $I \gets ImgGen(Img,l_x,l_y,S)$.
% \vspace{0.2em}
\STATE Calculate total classes: $T \gets Calclass(Img) $.

% \vspace{0.2em}
\FOR {Each class $n$ in $T$}
    \STATE Reset $Trigger~Counter$
    \FOR {Each trigger in $I$}
        \STATE Attach trigger:$Img_a= Combine(Img_n,trigger)$
        \STATE Calculate the highest output value $M_{max}$ and the corresponding class number $target~class$.
        \IF {$M_{max} < Th$}
            \STATE continue
        \ENDIF
        \IF {$target~class~!= n$}
            \STATE increment $Trigger~Counter$
        \ENDIF
        \IF {$counter > PO_{max}$}
            \STATE \textbf{return:} $P_1$
        \ENDIF
    \ENDFOR
\ENDFOR
\STATE Initialize trigger color: $C_1 \gets (R_1,G_1,B_1)$
\STATE Increment $Round Counter$
\IF{$Round Counter < CO_{max}$}
    \STATE Go to step 6
\ENDIF
% \vspace{0.2em}
\STATE \noindent \textbf{return} $P_2$
\end{algorithmic}
\end{algorithm}
\vspace{-1em}

\begin{figure}
    \centering
    \includegraphics[width=0.40\textwidth]{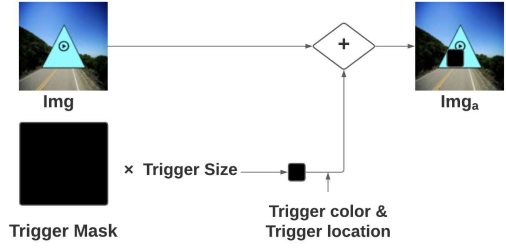} 
    \vspace{-1.6em}
    \caption{Example trigger generation in Algorithm~\ref{alg:polygon detection}.}
    \label{fig:demo}
    \vspace{-1em}
\end{figure}

Figure~\ref{fig:demo} shows the process to generate a $trigger$ and $Img_a$ in Algorithm~\ref{alg:polygon detection} with all possible trigger sizes. Given a trigger mask, we multiply it with $trigger~size$, and then apply $trigger~color$ to the trigger. \sys{} uses $trigger~location$ and obtains $Img_a$ by attaching the trigger on top of $Img$.

\vspace{-0.3em}
\subsection{Instagram Trojan Detection}
\vspace{-0.3em}
% \xq{add sample image. how does one filter work, explain some transformation}

To detect Instagram filter poisoned models, given the type of all used filters, we search each (source, target) class pair and apply each of the five filter types individually. These types are $GothamFilter$, $NashvilleFilter$, $KelvinFilter$, $LomoFilter$ and $ToasterFilter$. If the images in the source class have a high probability to be predicted as the target class after applying a certain filter, then the model is probably decided as Trojaned.

\vspace{-0.5em}
\begin{algorithm}[ht!]
\caption{Instagram Model Detection.}
\label{alg:ins detection}
\textbf{Input}: Model file($ID_n$); Clean images for each input class ($Img$).\\
\textbf{Parameter}: Threshold value to classify Instagram model ($Th$), Threshold count number ($Tc$), all the possible filter candidates ($S_{filters}$).

\textbf{Output}: A single probability that the model was poisoned ($P_1 = high~probability, P_2 = low~probability$).

\begin{algorithmic}[1]
\STATE Load model: $ID_n$.
\STATE Initialize trigger counter: $Counter$
\STATE Calculate total classes: $T \gets Calclass(Img)$.
\STATE Calculate total images of each class: $T_{img} \gets Calnumber(Img) $.

\FOR {Each class $n$ in $T$}
    \STATE Reset $counter$
    \FOR {Each image $Img_k$ in source class $n$}
        \FOR {Each Instagram filter type $s$ in all possible filter candidates $S_filter$}
            \STATE Obtain filtered image:$I \gets ImgGen(Img_k,s)$.
            \STATE Calculate the highest output value $M_{max}$ and the corresponding class number $target~class$.
            \IF {$M_{max} < Th$}
                \STATE continue
            \ENDIF
            \IF{$target~class~!= n$}
                \STATE increment $Counter$
            \ENDIF
            \IF{$counter >= T_{img} * Tc$}
                \STATE \textbf{return:} $P_1$
            \ENDIF
        \ENDFOR
    \ENDFOR
\ENDFOR
\STATE \noindent \textbf{return:} $P_2$
\end{algorithmic}
\end{algorithm}
\vspace{-1em}

Algorithm~\ref{alg:ins detection} shows the Instagram model detection method. The basic idea is to apple every candidate of filter type and check if the class has been changed. The threshold count number of this method is $1(100\%)$. If all candidates from one image flip their output class after applying a specific Instagram filter, the model is poisoned. Note that some clean images from benign models jump as well after applying an Instagram filter but not all of the images will jump. Therefore the threshold count number should be 1.

%% file: 5_evaluation.tex
\vspace{-1em}
\section{Evaluations}
\vspace{-0.3em}
% \huc{Why separated from Sec.6 ? we need to merge these two sections.}

\noindent \textbf{Experimental Setup.} Since the idea of this work comes from TrojAI~\cite{Trojai} Project and we use the data given by the competition to perform our preliminary experiment. Table~\ref{tab:dataset} shows a brief summary about the data. All of the models are trained by two different Adversarial Training approaches--Projected Gradient Descent (PGD) and Fast is Better than Free (FBF)~\cite{wong2020fast}. Note that we do not have access to the test data and holdout data. The performance is evaluated by uploading a container to a server and the server will evaluate the test data and holdout data by itself. The real-time result is posted on the TrojAI website. For train data, it is available to download form ~\cite{Trojai}. A detailed configuration of each model is shown in Table~\ref{tab:dataset}. 
We need to note that for each dataset, half of the models are poisoned. Polygon models and Instagram models take 50\% each of all poisoned models. 23 architectures are $resnet$, $wide\_resnet$, $densenet$, $googlenet$, $inceptionv3$, $squeezenetv$, $shufflenet$ and $vggs$ with different number of layers. The configureation of polygon trigger is shwon in Table~\ref{tab:dataset}. All candidates for Instagram triggers are $GothamFilter$, $NashvilleFilter$, $KelvinFilter$, $Toaster$, $LomoFilter$.

\vspace{-1em}
% https://pages.nist.gov/trojai/docs/data.html#round-3
\begin{table}[ht!]
\centering
\scalebox{0.76}{
\begin{tabular}{llll}
\hline
\multicolumn{1}{l}{} & \multicolumn{1}{l}{\begin{tabular}[c]{@{}l@{}}\# of models \end{tabular}} & \multicolumn{1}{l}{\begin{tabular}[c]{@{}l@{}}\# of model \\  architectures\end{tabular}} & Trigger type \\ \hline
CIFAR    & 10    & 1   & Polygon \& Instagarm \\ \hline
VGGface    & 10    & 1   & Polygon \& Instagarm \\ \hline
Train data   & 1008    & 23   & Polygon \& Instagarm \\ \hline
Test data    & 144     & 23   & Polygon \& Instagarm \\ \hline
Holdout data & 288     & 23   & Polygon \& Instagarm \\ \hline
\end{tabular}
}
\vspace{-0.6em}
\caption{TrojAI dataset.} 
% \huc{we can refer to metadata.csv file and report statistics such as trigger size.} 
\label{tab:dataset}
\end{table}
\vspace{-1.8em}

\begin{table}[ht!]
\centering
\begin{tabular}{lll} \hline
               & \begin{tabular}[c]{@{}l@{}}Polygon model \\ detection\end{tabular} & \begin{tabular}[c]{@{}l@{}}Instagram model \\ detection\end{tabular} \\ \hline
$Th $            & 0.999                                                             & 0.999                                                               \\ \hline
$PO_{max}$    & 3                                                                 & \textbackslash{}                                                    \\ \hline
$CO_{max}$    & 5                                                                 & \textbackslash{}                                                    \\ \hline
$Tc$             & \textbackslash{}                                          & 1                                                                   \\ \hline
$S_{filters}$ & \textbackslash{}                                          & 5 types \\ \hline                                                           
\end{tabular}
\vspace{-0.6em}
\caption{Parameter Configuration for model defections \label{tab:config}}
\vspace{-1em}
\end{table}

Table~\ref{tab:config} shows the configuration for both detection methods. The left column is for Polygon model detection and it has 3 parameters, the right column shows the parameters for Instagram model detection method.

\begin{figure}[ht!]
    \centering
    \begin{subfigure}[b]{0.49\columnwidth}
          \centering
          \includegraphics[width=\textwidth]{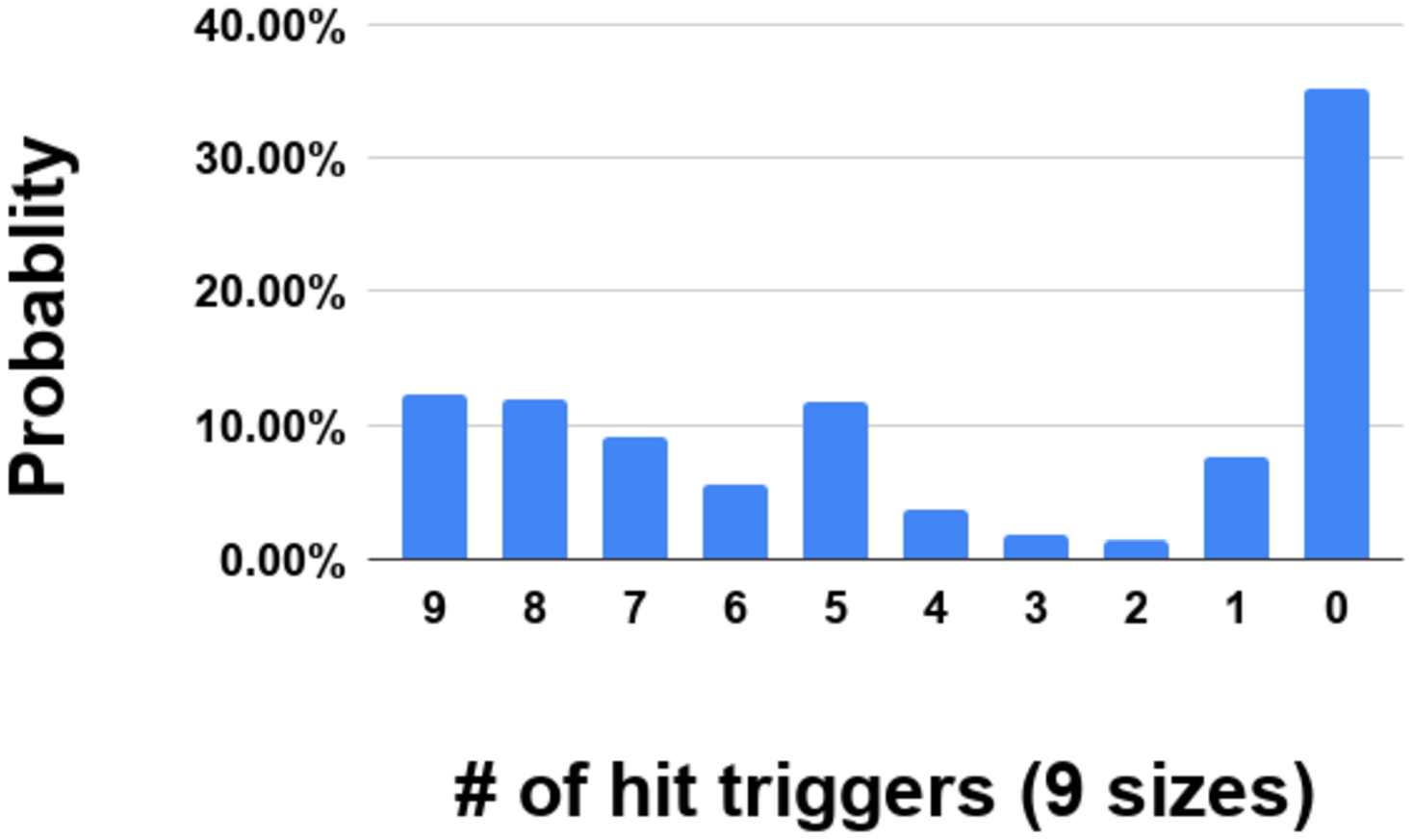}
          \caption{\label{fig:coverage_step}}
        %  \vspace{-0.35em}
    \end{subfigure}
    \hfill
    \begin{subfigure}[b]{0.49\columnwidth}
         \centering
         \includegraphics[width=\textwidth]{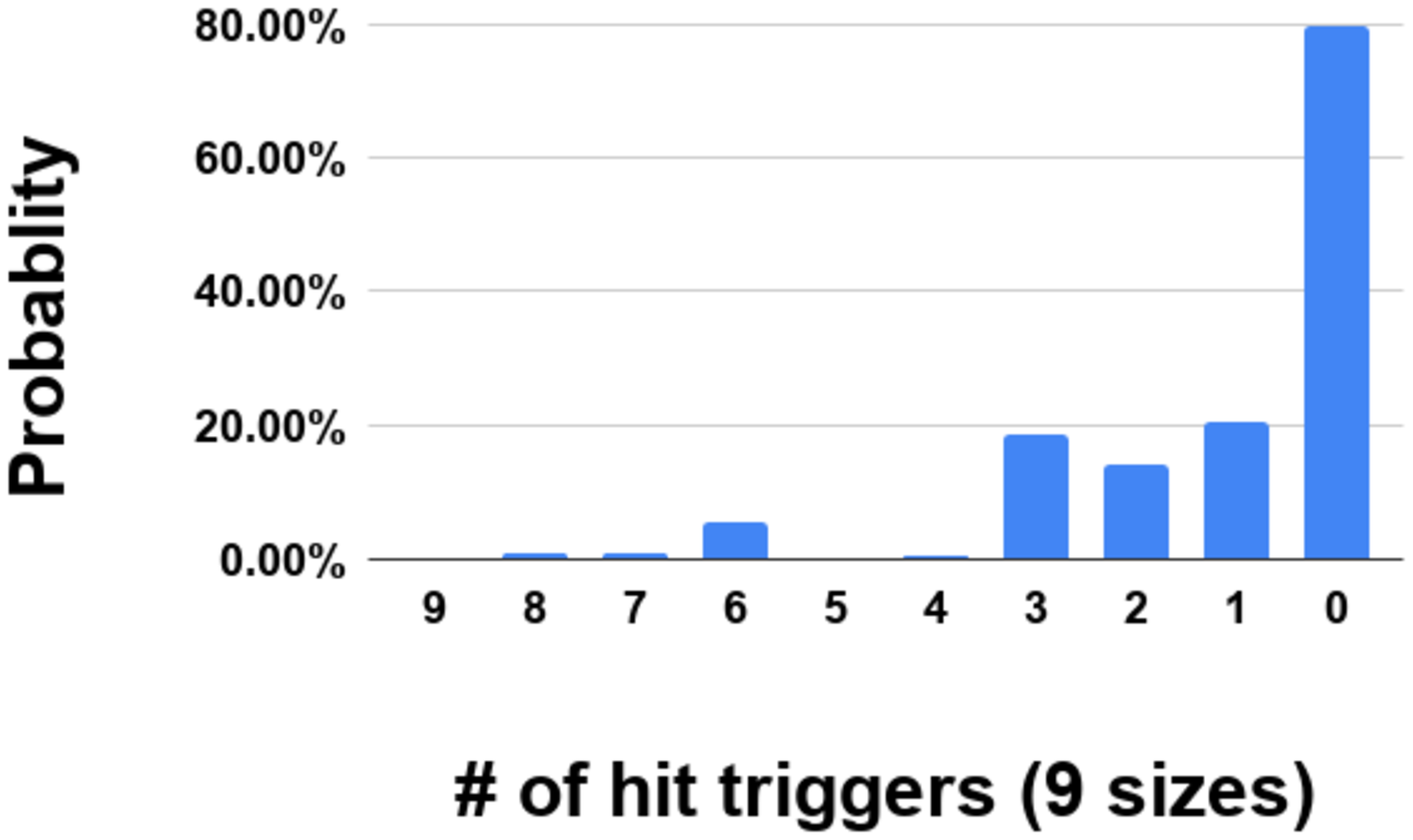}
         \caption{\label{fig:runtime_step}}
    \end{subfigure}
    \vspace{-0.7em}
    \caption{Trigger distribution.(a) poisoned model, (b)clean model.\label{fig:compare}}
\vspace{-1.5em}
\end{figure}

Figure~\ref{fig:compare} shows the trigger size distribution of the clean model and poisoned model. The x-axis represents the number of effective trigger sizes. There are total 9 trigger sizes and the poisoned model has a higher chance to have big number. We can see that the distribution is kind of difference between clean models and poisoned models.  A threshold of $3$ trigger sizes can offer best classification accuracy. The probability for detecting a poisoned model with a single random color is $54.14\%$. Cumulative probability measures the chance that one single catch happens during multiple independent events. In this case, 
% we use multiple independent random colors and find the cumulative probability of correctly classifying the poisoned models. 
We leverage the cumulative probability for different numbers of independent trigger colors. The error rate is defined as a model is falsely classified. 
% Based on some simple calculation, we know that
We find that as the number of random trigger colors increases, the error rate for poisoned model decreases and the error rate for the clean model increase a little. Therefore, we set the number $4$ to balance the total error rate.

\vspace{-1em}

We evaluate our method on the self-generated CIFAR, VGGFace dateset and the TrojAI round-3 train, test and holdout dataset. 
The CrossEntropyLoss$(CE-Loss)$ is defined as:
\vspace{-0.3em}
\begin{equation}
    CrossEntropyLoss = - (y \cdot log(p) +(1 - y) \cdot log (1-p)
\end{equation}

Table~\ref{tab:train_eva_data} shows the result of both train data and test data. We can see consistent performance without overfitting problem. 

% Since the AUC score is only slightly better than the target AUC score(0.90), our method still has a lot of potentials to explore and optimize. 

After that, we further evaluate our method for holdout data. Table~\ref{tab:hold_data} shows the evaluation results. Note that we find out false prediction are mostly false positives and half of the false positives are detected as Polygon model while the other half are detected as Ins model. This is a very interesting situation and we need to deal with the model individually. Therefore We have been trying different methods to minimize these false positives. The method is very tricky because the false plosives have almost identical responses comparing with Trojaned models. We suspect these models are "fake" clean models and we still need to dig into it to find the hidden difference between real Trojaned models and "fake" clean models. 

\vspace{-0.6em}
\begin{table}[ht!]
\centering
\scalebox{0.76}{
\begin{tabular}{llll}
\hline
Models & \multicolumn{1}{l}{CE-Loss} & \multicolumn{1}{l}{ROC-AUC} & \multicolumn{1}{l}{Runtime(s)} \\ \hline
CIFAR & 0.00 & 1.00 & 47.4  \\ \hline
VGGFace & 0.00 & 1.00 & 119.6  \\ \hline
Train Models  & \multicolumn{1}{l}{} & \multicolumn{1}{l}{} & \multicolumn{1}{l}{} \\ \hline
Id-00000000-099 & 0.3464 & 0.890 & 64696  \\ \hline
Id-00000100-199 & 0.3254 & 0.900  & 112574 \\ \hline
100 Random models  & 0.3254 & 0.900  & 40688  \\ \hline 
Test Models & \multicolumn{1}{l}{} & \multicolumn{1}{l}{} & \multicolumn{1}{l}{} \\ \hline
144 Test models & 0.3113 & 0.906 & 122400 \\ \hline
\end{tabular}
}
\vspace{-0.6em}
\caption{\label{tab:train_eva_data}Evaluation on different data.}
\end{table}

\vspace{-1.6em}
\begin{table}[ht!]
\centering
\scalebox{0.82}{
\begin{tabular}{llll}
\hline
Models & \multicolumn{1}{l}{CE-Loss} & \multicolumn{1}{l}{ROC-AUC} \\ \hline
144 clean models & 0.3647 & 0.8819 \\ \hline
72 Polygon models & 0.3495 & 0.8889 \\ \hline
72 Instagram models & 0.2884 & 0.9167 \\ \hline
Total 288 models & 0.3419 & 0.8924 \\ \hline
\end{tabular}
}
\vspace{-0.5em}
\caption{\label{tab:hold_data}Evaluation on holdout data.}
\end{table}
\vspace{-1em}

Even though our method enables very high performance, it still has some drawbacks. For example, there is a very small $0.02$ difference of ROC-AUC because of the randomness of color initialization, which is acceptable. Our future direction focus on improving the consistent output accuracy and runtime. We will also introduce text trigger into our method.

%% file: 6_conclusion.tex
\vspace{-1em}
\section{Conclusion}
\vspace{-0.6em}
In this paper, We propose the first trigger approximation based Trojan detection framework \sys{}. It enables a fast and scalable search of the trigger in the input space.
Also, \sys{} can also detect Trojans embedded in the feature space where certain filter transformations are used to activate the Trojan. Empirical results show that TAD achieves a superior performance.